\documentclass[twocolumn,showpacs,preprintnumbers,amsmath,amssymb,prb]{revtex4}

\usepackage{graphicx}% Include figure files
\usepackage{dcolumn}% Align table columns on decimal point
\usepackage{bm}% bold math
\usepackage{hyperref}

\begin{document}

\title{
Energy dissipation in DC-field driven electron lattice 
coupled to fermion baths
}

\author{Jong E. Han and Jiajun Li}
\affiliation{
Department of Physics, State University of New York at Buffalo, Buffalo,
New York 14260, USA}

\date{\today}

\begin{abstract} 

Electron transport in electric-field-driven tight-binding lattice
coupled to fermion baths is comprehensively studied. We reformulate the
problem by using the scattering state method within the Coulomb
gauge.  Calculations show that the formulation justifies direct access
to the steady-state bypassing the time-transient calculations, which
then makes
the steady-state methods developed for quantum dot theories applicable
to lattice models. We show that the effective temperature of the
hot-electron induced by a DC electric field behaves as $T_{\rm
eff}=C\gamma(\Omega/\Gamma)$ with a numerical constant $C$,
tight-binding parameter $\gamma$, the Bloch oscillation frequency
$\Omega$ and the damping parameter $\Gamma$. In the small damping limit
$\Gamma/\Omega\to 0$, the steady-state has a singular property with the
electron becoming extremely hot in an analogy to the short-circuit
effect. This leads to the conclusion that the dissipation mechanism
cannot be considered as an implicit process, as treated in equilibrium
theories. Finally, using the energy flux relation, we derive a
steady-state current for interacting models where only on-site Green's
functions are necessary.

\end{abstract}


\maketitle

\section{Introduction}

Formulating strong-field transport in electron lattice
has always been one of the most challenging
theoretical goals in condensed matter physics~\cite{kadanoff,mahan}. This is more true with
today's advanced nano-lithography techniques where we can now realize
electron lattice driven far from equilibrium. Even though this is an old
problem, we are still in search of a firm theoretical paradigm to approach the
problem in general. One of the central puzzles is dissipation. In equilibrium, the
presence of an open environment in contact with a system introduces
thermalization, and once the temperature is defined, we often use
canonical or grand canonical ensemble, and do not consider the coupling
to the environment as an explicit parameter. We naturally ask whether
such simplifying ansatz may be possible in nonequilibrium, at least for
steady-state description.

In nonequilibrium, we do not know such tremendously simplifying
paradigms to take the role of the environment as an implicit parameter
which can be hidden in thermalization process. On the contrary, the
dissipation is considered as an integral part of the
nonequilibrium process, and we need to include the dissipation mechanism
explicitly for sound theoretical description. In some systems, however,
the dissipation process can be simplified. In quantum dots (QDs) under a
finite DC bias, electron reservoirs coupled to the QD also act as energy
source/drain and the energy relaxation is assumed to happen far away from QD and
inside a battery. The electrical leads are then modeled as
non-interacting reservoirs,
and the electron transport is viewed as a transmission
problem~\cite{landauer}.  By taking the open limit, the excess energy
can be taken infinitely far away from the QD, and the problem supports
steady-state~\cite{doyon}. Various quantum simulation methods have been
proposed to study the transient behaviors of interacting
models~\cite{tdnrg,werner,schiro}. In the limit where a steady-state
exists, the nonequilibrium state can also be described within the time-independent
statistical mechanics framework~\cite{zubarev}, from which
Hershfield~\cite{hershfield} proposed the nonequilibrium density matrix 
\begin{equation}
\hat{\rho}=\exp\left[-\beta\sum_k\sum_{\alpha=S,D} (\epsilon_{\alpha
k}-\mu_\alpha)\psi^\dagger_{\alpha k}\psi_{\alpha k} \right],
\end{equation} 
with the reservoir energy $\epsilon_{\alpha k}$ for the source
($\alpha=S$) and drain ($D$) with the continuum index $k$.
$\psi^\dagger_{\alpha k}$ is the creation operator of the full
scattering state as the solution of the \textit{whole} system of quantum
dot and the leads.
$\mu_\alpha$ is the chemical potential of the respective reservoirs.
The scattering state formulation, although conceptually appealing, has
initially been applied only to limited models~\cite{schiller} due to the
difficulty of finding the scattering states. In recent years, several
steady-state methods~\cite{mehta,prl07,anders,prb06} have been developed and have
been extended to general models. 

Recently, the attention of the field has turned to lattice nonequilbrium
problems. Even at a very stage of the field, there have been important
findings in the nonequilbrium processes, most notably that electrons
under a DC electric field seem to build up internal energy quite
quickly, reaching a quite different steady-state from equilibrium
strongly correlated states~\cite{freericks}.  One of the most popular
technique of solving lattice many-body models has been the dynamical
mean-field theory (DMFT)~\cite{dmft}. Its success has been well
documented in the description of the Mott transition and
strong-correlation physics in equilibrium. While there have been a fair
amount of DMFT works to electric-field driven lattice systems, the
validity of the local approximation is still unconfirmed and subject to
intense debate.

There have been numerous attempts to simulate nonequilibrium physics in
lattice models, often through isolated
Hamiltonians~\cite{turkowski,freericks,eckstein,aoki} suited for quench
dynamics of cold atom systems in optical lattice, periodically driven
systems~\cite{aoki,demler}, and some basic
dissipation models~\cite{aoki,amaricci,aron,aron2,vidmar}.  However,
in part due to the numerical difficulties in simulating long time-evolution,
most of the efforts have focused on high-field phenomena such as the
dielectric breakdown in Mott insulators~\cite{eckstein,oka}. The main emerging picture of
the calculations is that the external field drives the electronic
lattice systems into hot temperature, generally regardless of the nature
of many-body interaction. Even though the picture is in agreement
between many groups, the detailed understanding of the nature of the hot
electron state and its eventual fate in more realistic setting is not
available. To gain systematic understanding of such nonequilibrium
state, the dissipation should be included in explicit models and their
analytic behavior has to be studied with the damping as a controlled
parameter.

One of the goals of this paper is to introduce steady-state formulation
via scattering state method for lattice with fermion baths and
comprehensively analyze the model to show that the system possesses many
properties which are expected of physical systems, for instance,
consistent picture as the Boltzmann transport theory. In the process, an
argument will be made that the fermion bath model and the steady-state
methods are a good minimal system to study nonequilibrium strong
correlation physics.  In the previous paper~\cite{fbath} by one of
Authors, the fermion bath model under a DC electric field has been shown
to reproduce the key ingredients as predicted by the classical Boltzmann
transport theory, and to have a stead-state solution. The occupation
number as a function of mechanical momentum has been shown to have the
Fermi sea shift by the drift velocity proportional to the scattering
time and the electric field. Furthermore, the DC current has been
derived to be consistent with the Boltzmann transport result applied to
nanostructures~\cite{lebwohl}.

In this work, we further develop the solution to show that the
scattering state formulation is applicable, and therefore a wide range of
new techniques can be developed to solve the interacting lattice
nonequilibrium phenomena. Explicit calculations from temporal gauge and
the Coulomb gauge with scattering state formulation show that they are
completely consistent with each other. The Coulomb gauge enables the
time-independent formalism, making physical interpretations more
transparent. We calculate explicitly the local distribution
function, from which we derive that the effective temperature scales as
$T_{\rm eff}=C\gamma(\Omega/\Gamma)$ with a numerical constant $C$,
tight-binding parameter $\gamma$, the Bloch oscillation frequency
$\Omega$ and the damping parameter $\Gamma$. The effective temperature
exhibits a singular limit of $T_{\rm eff}\to\infty$
for $\Gamma/\Omega\to 0$. This proves that one should not take the
damping as an implicit process, as treated in equilibrium theory.
Finally we derive, via the energy flux conservation with the Joule
heating, a general DC current relation as a functional of local Green's
functions as an extension of the Meir-Wingreen formula~\cite{meir} to lattice
models, and confirm the linear response theory.

The main text of the paper is organized as follows. In Section II, the
method introduced in Ref.~\onlinecite{fbath} is further developed for
Green's functions in the temporal gauge. In Section III, we introduce
the Coulomb gauge and show that the Green's functions in both gauges
become identical in the long-time limit. In Section III we further  discuss several
important nonequilibrium quantities: the local
distribution function and the effective temperature
in A, time-evolution of wave-packet in B, dissipation and energy flux in
C, and finally the derivation of the DC current in interacting models in D.
Appendices provide detailed analytic calculations.

\section{Time-dependent theory with temporal gauge}
\begin{figure}
\rotatebox{0}{\resizebox{2.2in}{!}{\includegraphics{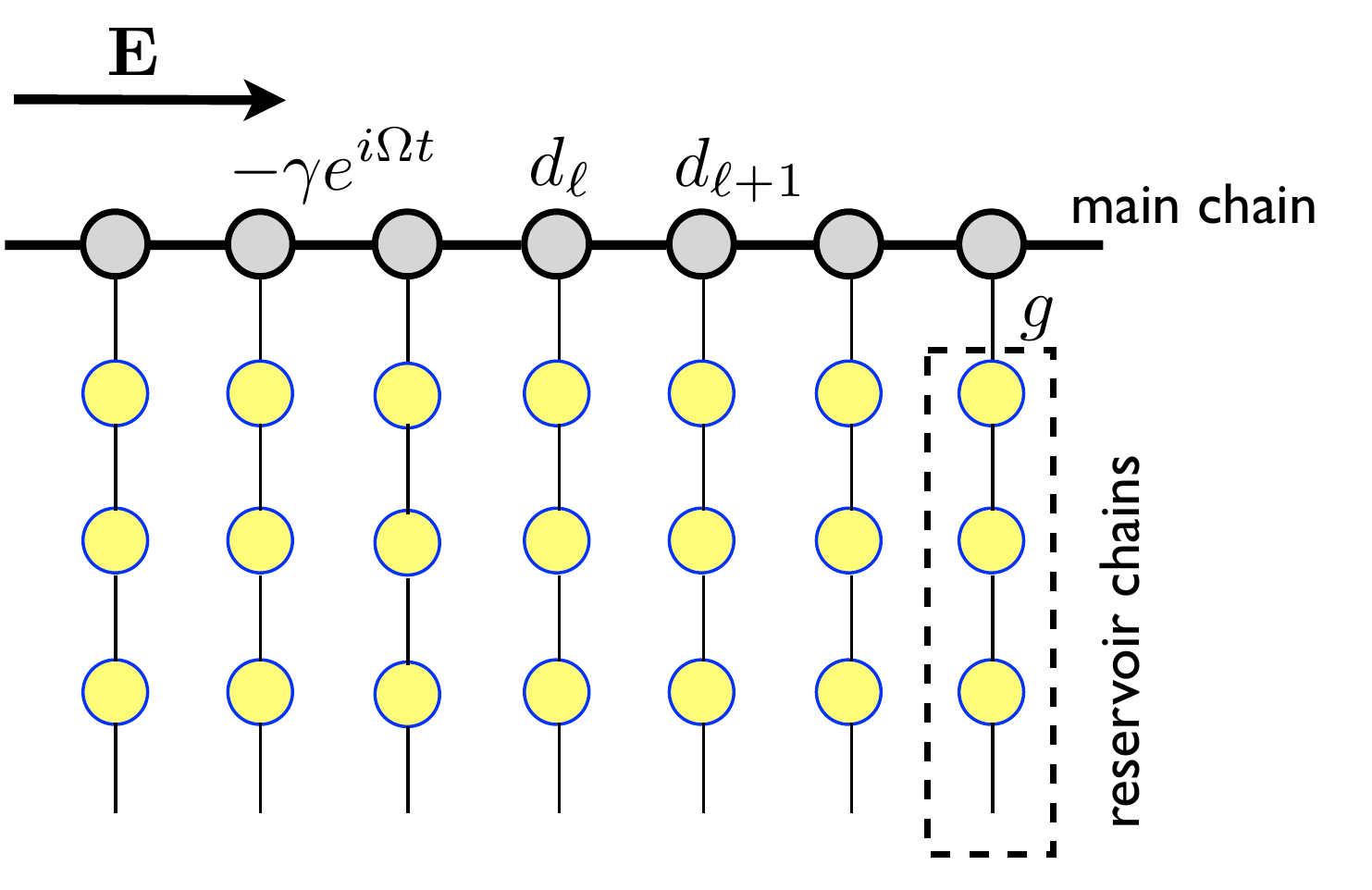}}}
\caption{One-dimensional tight-binding lattice of orbital $d_\ell$ under
an electric field $E$. Each lattice site is connected to an identical
fermionic bath of $\{c_{\ell\alpha}\}$ with the continuum index $\alpha$ along the
reservoir chain direction. In the temporal gauge the effect of the
electric field is absorbed in the hopping integral $-\gamma$ with the
Peierls phase $e^{i\Omega t}$.
}
\label{fig1}
\end{figure}

To demonstrate the equivalence of the time-dependent temporal gauge to
the scattering-state formalism with time-independent Coulomb gauge, we
start with the one-dimensional non-interacting model considered
earlier~\cite{fbath}.
We study a quadratic model of a one-dimensional $s$-orbital
tight-binding model connected to fermionic reservoirs (see
Fig.~\ref{fig1}) under a uniform electric field $E$. The effect of the electric
field for time $t>0$ is absorbed in the temporal gauge as the Peierls phase
$\varphi(t)=\Omega t \cdot\theta(t)$ in the
hopping integral~\cite{turkowski} $\gamma$. Here 
$\Omega=eEa$ is the Bloch oscillation frequency and $\theta(t)$ is the step
function. The time-dependent Hamiltonian then reads
\begin{eqnarray}
\hat{H}(t)&=&-\gamma\sum_\ell (e^{i\varphi(t)}
d^\dagger_{\ell+1}d_\ell+{\rm H.c.})+\sum_{\ell\alpha}\epsilon_\alpha
c^\dagger_{\ell\alpha}c_{\ell\alpha}\nonumber \\
&  &-\frac{g}{\sqrt{L}}\sum_{\ell\alpha}(c^\dagger_{\ell\alpha}d_\ell+{\rm H.c.}),
\end{eqnarray}
with $d^\dagger_\ell$ as the (spinless) electron operator on the
tight-binding chain on site $\ell$, $c^\dagger_{\ell\alpha}$ with the
reservoir fermion states connected to the site $\ell$ with the continuum index
$\alpha$ along each reservoir chain of length $L$. The length $L$ is
taken to infinity, and the time scale $L/v_F$ (with Fermi velocity of
the chain $v_F$) for the wave to reach the end of the reservoir chain is
considered larger than any other time scales in the
problem. As discussed in
Ref.~\onlinecite{fbath}, the Hamiltonian can be diagonalized in each
$k$-sector as $\hat{H}(t)=\sum_k \hat{H}_k(t)$ with
\begin{eqnarray}
\hat{H}_k(t)&=&
-2\gamma\cos[k+\varphi(t)]d^\dagger_kd_k+\sum_{\alpha}\epsilon_\alpha
c^\dagger_{k\alpha}c_{k\alpha}\nonumber \\
&& -\frac{g}{\sqrt{L}}
\sum_{\alpha}(c^\dagger_{k\alpha}d_k+{\rm H.c.}),
\end{eqnarray}
with the fermion operators Fourier transformed to the wave-vector basis.
Here $\epsilon_{d}(k)=-2\gamma\cos(k)$ is the tight-binding dispersion
at zero $E$-field. The reservoir states formed by $c^\dagger_{k\alpha}$
acts as an open particle source with its chemical potential set at zero
energy. The problem can be solved with $\hat{H}_0=\hat{H}_k(0)$ as the unperturbed
Hamiltonian and $\hat{V}(t)=\hat{H}_k(t)-\hat{H}_k(0)$ as the time-dependent
perturbation,
\begin{equation}
\hat{V}(t) = -2\gamma\left\{\cos[k+\varphi(t)]-\cos(k)\right\}d^\dagger_k
d_k \equiv v(t)d^\dagger_k d_k.\label{eq:pert}
\end{equation}
This block-diagonal Hamiltonian is nothing but a resonant level model
coupled to a reservoir, with the level's energy oscillating in
time~\cite{jauho}.
With the perturbation of one-body terms of a finite degrees of freedom,
one can write the Dyson's equation for the retarded and lesser Green's
functions as~\cite{blandin,fbath}
\begin{eqnarray}
{\bf G}_k^r&=&{\bf G}_{k,0}^r+{\bf G}_{k,0}^r{\bf V}{\bf G}_k^r
\label{eq:gr}\\
{\bf G}_k^<&=&[I+{\bf G}_k^r{\bf V}]{\bf G}_{k,0}^<[I+{\bf V}{\bf
G}_k^a],\label{eq:glss}
\end{eqnarray}
where ${\bf G}_k^<$ and ${\bf G}_k^r$ are the lesser and retarded
Green's function matrices, respectively. ${\bf G}_{k,0}^<$ and ${\bf
G}_{k,0}^r$ are for the non-interacting limit.
The multiplication of Green's function matrices denotes time
integration. 

Following Ref.~\onlinecite{fbath}, the retarded Green's function is
\begin{equation}
G_k^r(t,t')=-i\theta(t-t')
e^{-\Gamma|t-t'|+2\gamma i\int_{t'}^t \cos(k+\Omega s)ds},
\end{equation}
with the damping parameter $\Gamma=(g^2/L)\sum_\alpha
\delta(\epsilon_\alpha)$ for reservoirs of flat density of states of
infinite bandwidth.
The local retarded Green's function $G^r_{\rm
loc}(t,t')=(2\pi)^{-1}\int^\pi_{-\pi}G^r_k(t,t')dk$ becomes
\begin{equation}
G_{\rm loc}^r(t,t')=-i\theta(t-t')e^{-\Gamma|t-t'|}J_0\left(
\frac{4\gamma}{\Omega}\sin\frac{\Omega(t-t')}{2}
\right),
\end{equation}
with the zero-th Bessel function $J_0(x)$. Here, the gauge-invariant
local function
becomes a function of only the relative time, $G_{\rm loc}^r(t,t')=G_{\rm
loc}^r(t-t')$. Fourier transformation with respect to the relative time
gives
\begin{equation}
G^r_{\rm
loc}(\omega)=\sum_{m=-\infty}^\infty\frac{J_m(\frac{2\gamma}{\Omega})^2}{
\omega+m\Omega+i\Gamma},
\label{eq:gtr}
\end{equation}
by using the Bessel function relation~\cite{gradshteyn} $J_0(2z\sin\frac{\alpha}{2})=\sum_m
[J_m(z)]^2e^{im\alpha}$.

The lesser Green's function can be simplified in a straightforward
calculation from Eq.~(\ref{eq:glss}) following the similar procedures as in
Ref.~\onlinecite{fbath} in the long-time limit ($t,t'\gg\Gamma^{-1}$) as
\begin{equation}
G_k^<(t,t')=\int^t_{-\infty}ds\int^{t'}_{-\infty}ds'
G^r_k(t,s)\Sigma_\Gamma^<(s-s')G^a_k(s',t'),
\end{equation}
where 
\begin{equation}
\Sigma_\Gamma^<(s)=\int^0_{-\infty}
\frac{i\Gamma}{\pi}e^{-i\omega s}d\omega,
\end{equation}
with the self-energy ${\bf \Sigma}_\Gamma$ from the damping taken as the
perturbation. Although the above equation has been derived with the
time-dependent Peierls term
Eq.~(\ref{eq:pert}) as the perturbation, the same result can be obtained when the damping is
considered as perturbation in the limit that transient terms die out.
The local lesser Green's function can be computed as $G^<_{\rm
loc}(t,t')=(2\pi)^{-1}\int_{-\pi}^\pi G^<_k(t,t')dk$, which again renders the
Green's function only dependent on the relative time. After changing the
dummy variables $s-t\to s$ and $s'-t'\to s'$, we have
\begin{eqnarray}
G^<_{\rm loc}(\omega) & = &
\frac{i\Gamma}{\pi}\int_{-\infty}^\infty dt\int_{-\infty}^0d\omega'\int_{-\infty}^0ds
\int_{-\infty}^0ds' \nonumber \\
& \times & e^{i(\omega-\omega')t-i\omega'(s-s')+\Gamma(s+s')}
J_0\left(\frac{4\gamma}{\Omega}\sqrt{A}\right)\nonumber
\end{eqnarray}
with
\begin{eqnarray}
A & = & \sin^2\frac{\Omega s}{2}+\sin^2\frac{\Omega
s'}{2}-2\cos\left[\Omega\left(t+\frac{s-s'}{2}\right)\right]\nonumber \\
& & \times
\sin\frac{\Omega s}{2}\sin\frac{\Omega s'}{2}.\nonumber
\end{eqnarray}
Again by utilizing the Bessel function relation~\cite{gradshteyn}
$J_0(\sqrt{a^2+b^2-2ab\cos\alpha})=\sum_mJ_m(a)J_m(b)e^{im\alpha}$
\begin{eqnarray}
G^<_{\rm loc}(\omega)  & = &
2i\Gamma\sum_mf(\omega+m\Omega)
\left|
\int_{-\infty}^0 e^{-i(\omega+\frac{m\Omega}{2})s+\Gamma s}\right.
\nonumber \\
& & \times\left.
J_m\left(\frac{4\gamma}{\Omega}\sin\frac{\Omega s}{2}
\right)ds
\right|^2 ,\nonumber
\end{eqnarray}
with $f(x)=\theta(-x)$, the Fermi-Dirac function at zero temperature.
From the identity~\cite{gradshteyn},
$J_m(2z\sin\frac{\alpha}{2})=e^{-im(\pi-\alpha)/2}\sum_\ell
J_\ell(z)J_{m+\ell}(z)e^{i\ell\alpha}$,
\begin{equation}
G^<_{\rm loc}(\omega)  = 
2i\Gamma\sum_mf(\omega+m\Omega)
\left|\sum_\ell\frac{J_\ell(\frac{2\gamma}{\Omega})
J_{\ell-m}(\frac{2\gamma}{\Omega})}{\omega+\ell\Omega+i\Gamma}
\right|^2 .
\label{eq:gtlss}
\end{equation}
Although the two Green's functions, Eqs.~(\ref{eq:gtr}) and
(\ref{eq:gtlss}), have been reduced to familiar forms
resembling spectral representation, a clear relation between them is not
available yet. In the following section, we discuss the scattering state
formalism and find relations connecting the retarded and lesser Green's
functions.

\section{Scattering state formalism}

\begin{figure}
\rotatebox{0}{\resizebox{2.4in}{!}{\includegraphics{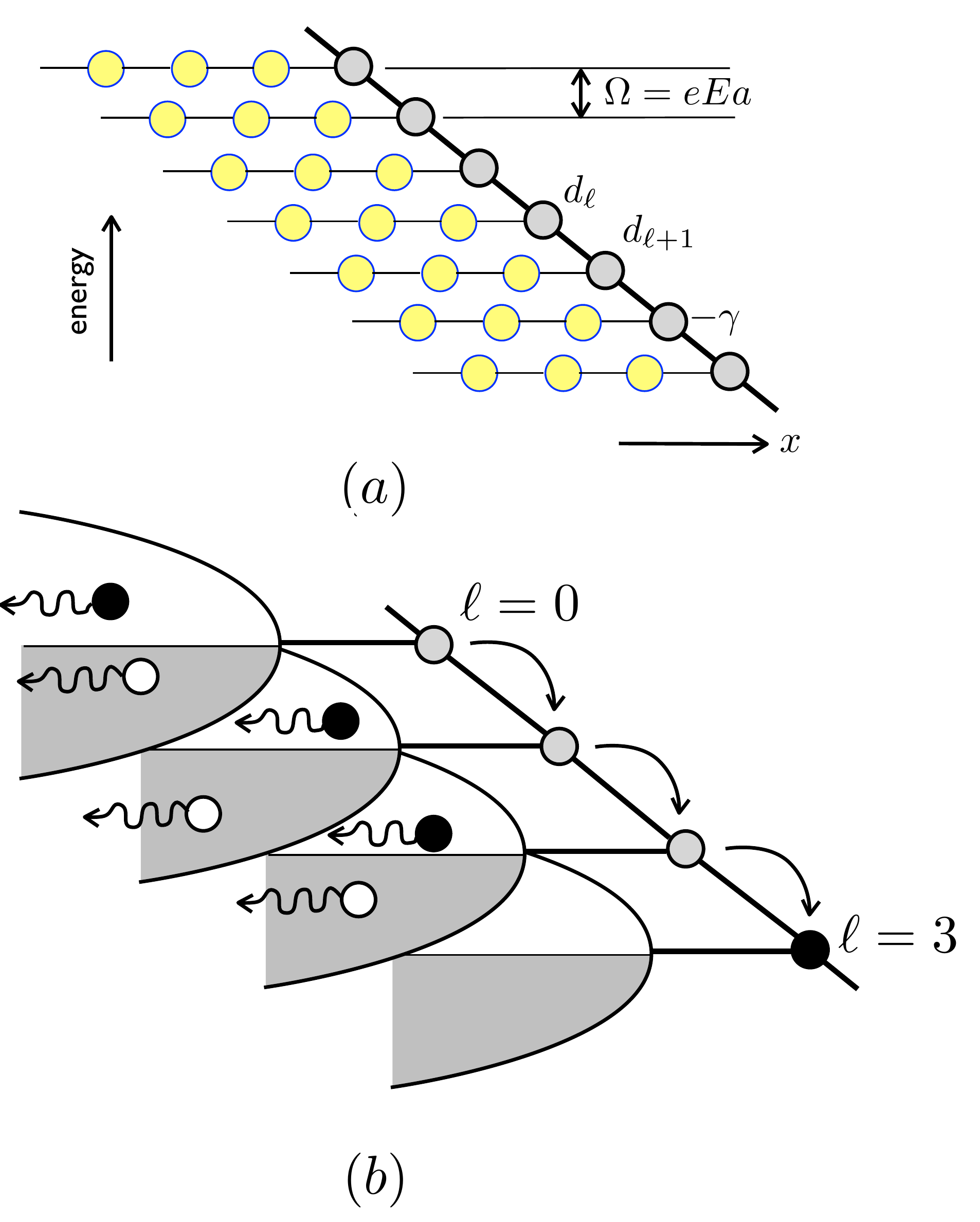}}}
\caption{
(a) In the Coulomb gauge for the scattering state formulation, each
tight-binding sites and the
associated fermion baths are on a potential slope with the potential
drop between neighboring sites as $\Omega$.
(b) As electron moves down the potential slope (from $\ell=0$ to
$\ell=3$ in the figure), it leaves a trail of electron-hole pairs in the
fermion reservoirs through particle-exchange. Since these electron-hole pairs travel indefinitely
along the infinite bath chains, the reservoirs act like energy
drains mimicking inelastic processes. There is an energy flux to the
reservoirs, but no net particle flux.
}
\label{fig2}
\end{figure}

We have learned from Ref.~\onlinecite{fbath} and the above calculations
that the fermion bath model has a well-defined time-independent limit for
gauge-invariant quantities such as local Green's function. This
observation and the presence of infinite reservoir states prompt us to
consider scattering state formalism~\cite{scattering}. As depicted in Figs.~\ref{fig2}(a),
for each site on the main chain there are infinite degrees of freedom
coupled from each reservoirs. Therefore, we can rewrite the quadratic
Hamiltonian in terms of the scattering states originating from the
fermion reservoir chains~\cite{prb06}. To adopt the time-independent scattering
theory, we use the Coulomb gauge as shown in Fig.~\ref{fig2}(a) with the
Hamiltonian,
\begin{eqnarray}
\hat{H}_{\rm Coul}&=&-\gamma\sum_\ell (
d^\dagger_{\ell+1}d_\ell+{\rm H.c.})
-\sum_\ell \ell\Omega d^\dagger_\ell d_\ell
\nonumber \\ 
& & +\sum_{\ell\alpha}(\epsilon_\alpha-\ell\Omega)
c^\dagger_{\ell\alpha}c_{\ell\alpha}
\label{eq:hcoul}\\
& & -\frac{g}{\sqrt{L}}\sum_{\ell\alpha}(c^\dagger_{\ell\alpha}d_\ell+{\rm
H.c.}),\nonumber
\end{eqnarray}
where the static Coulomb potential is applied as a potential slope to
the chain and each reservoirs with their chemical potential are raised
together with the corresponding
tight-binding sites.

Since the Hamiltonian is quadratic and the chain is coupled to open
systems, we can decompose the Hamiltonian in terms of the scattering
state operators $\psi^\dagger_{\ell\alpha}$ originating from the asymptotic
state $c^\dagger_{\ell\alpha}$ as given by the Lippmann-Schwinger
equation~\cite{scattering},
\begin{eqnarray}
\psi^\dagger_{\ell\alpha} & = & c^\dagger_{\ell\alpha}
+\frac{1}{\epsilon_\alpha-\ell\Omega-{\cal
L}+i\eta}[\hat{H}_g,c^\dagger_{\ell\alpha}]\nonumber \\
& = &
c^\dagger_{\ell\alpha}
-\frac{g}{\sqrt{L}}\frac{1}{\epsilon_\alpha-\ell\Omega-{\cal 
L}+i\eta}d^\dagger_{\ell},
\end{eqnarray}
with the Liouvillian operator ${\cal L}\hat{A}=[\hat{H}_{\rm Coul},\hat{A}]$ for
any operator and
$\hat{H}_g=-(g/\sqrt{L})\sum_{\ell\alpha}(c^\dagger_{\ell\alpha}d_\ell+{\rm
H.c.})$. The scattering state embodies the openness of the fermion
reservoirs. As implemented by the infinitesimal imaginary poles given by $i\eta$,
once the electrons scatter into a reservoir they \textit{never} come back
to the tight-binding chain.
With the scattering state basis, the Hamiltonian is rewritten as
\begin{equation}
\hat{H}_{\rm
Coul}=\sum_{\ell\alpha}(\epsilon_\alpha-\ell\Omega)\psi^\dagger_{\ell\alpha}
\psi_{\ell\alpha},
\end{equation}
with the Fermi statistics applied separately within each $\ell$-sector
\begin{equation}
\langle \psi^\dagger_{\ell\alpha}\psi_{\ell\alpha}\rangle =
f(\epsilon_\alpha).
\end{equation}

Since the Hamiltonian is quadratic, the scattering state operator
$\psi^\dagger_{\ell\alpha}$ is a linear combination of the original
basis of $d^\dagger_{\ell'}$ and $c^\dagger_{\ell'\alpha'}$, and
\begin{equation}
\psi^\dagger_{\ell\alpha} = c^\dagger_{\ell\alpha}
+\sum_{\ell'}d^\dagger_{\ell'}C_{\ell\alpha}(d_{\ell'})
+\sum_{\ell'\alpha'}c^\dagger_{\ell'\alpha'}C_{\ell\alpha}(c_{\ell'\alpha'}),
\end{equation}
with the expansion coefficient for a fermion annihilation operator $a$ given as
\begin{equation}
C_{\ell\alpha}(a)=\{a,\psi^\dagger_{\ell\alpha}-c^\dagger_{\ell\alpha}\}=
\left\{
a,\frac{-g/\sqrt{L}}{\epsilon_\alpha-\ell\Omega-{\cal
L}+i\eta}d^\dagger_\ell
\right\}.
\end{equation}
It is important to realize that, for a quadratic Hamiltonian, the
anti-commutation of operators in $C_{\ell\alpha}(a)$ is just a
c-number. In fact, this c-number is nothing but the retarded Green's function
$\overline{G}^r_{\ell'\ell}(\epsilon_\alpha-\ell\Omega)$ between
$d_{\ell'}$ and $d^\dagger_\ell$. From this argument, the retarded
Green's function in any time-independent quadratic Hamiltonian is
\textit{independent of statistics}. Note that the scattering state
$\psi^\dagger_{\ell\alpha}$
originating from the $\ell$-th reservoir has admixture from any
reservoir states $c^\dagger_{\ell'\alpha'}$. Here, we have put an over-line to
denote the Green's function in the Coulomb gauge. We then have
\begin{equation}
\psi^\dagger_{\ell\alpha}  =  c^\dagger_{\ell\alpha}
-\frac{g}{\sqrt{L}}\sum_{\ell'}\overline{G}^r_{\ell'\ell}(\epsilon_\alpha-\ell\Omega)
d^\dagger_{\ell'}+\cdots.
\end{equation}
This equation can be inverted to express $d_\ell$ in terms of the scattering
state basis $\psi_{\ell\alpha'}$ as
\begin{equation}
d_{\ell}  =
\sum_{\ell'\alpha'}\tilde{C}_\ell(\ell'\alpha')\psi_{\ell'\alpha'}.
\end{equation}
Similarly as above,
$\tilde{C}_\ell(\ell'\alpha')=\{\psi^\dagger_{\ell'\alpha'},d_\ell\}$
since the scattering state operators also satisfy the anti-commutation
relation~\cite{prb06}
$\{\psi^\dagger_{\ell,\alpha},\psi_{\ell'\alpha'}\}=\delta_{\ell\ell'}\delta_{\alpha\alpha'}$,
and
\begin{equation}
d_{\ell}  =  
-\frac{g}{\sqrt{L}}\sum_{\ell'\alpha'}\overline{G}^r_{\ell\ell'}(\epsilon_{\alpha'}-\ell'\Omega)
\psi_{\ell'\alpha'}.
\end{equation}
We will discuss below how $\overline{G}^r_{\ell\ell'}(\omega)$ is
explicitly calculated. On-site retarded Green's function at the central
site $\ell=0$ can be expressed in terms of the scattering
state basis as
\begin{eqnarray}
\overline{G}^r_{00}(\omega)&=&\frac{g^2}{L}\sum_{\ell\alpha,\ell'\alpha'}
\frac{\overline{G}^r_{0\ell}(\epsilon_\alpha-\ell\Omega)
[\overline{G}^r_{0\ell'}(\epsilon_{\alpha'}-\ell'\Omega)]^*}{\omega-\epsilon_\alpha+\ell\Omega
+i\eta}\nonumber \\
& &\times\langle\{\psi_{\ell\alpha},\psi^\dagger_{\ell'\alpha'}\}\rangle
\nonumber \\
& = & \frac{g^2}{L}\sum_{\ell\alpha}
\frac{|\overline{G}^r_{0\ell}(\epsilon_\alpha-\ell\Omega)|^2}{
\omega-\epsilon_\alpha+\ell\Omega+i\eta}.
\label{eq:gwr}
\end{eqnarray}
Here the retarded Green's functions appear on both sides of the
equation, and its self-consistency will be examined below.

The lesser Green's function can be calculated similarly as
\begin{eqnarray}
\overline{G}^<_{\ell\ell'}(\omega)
& = & i\frac{2\pi g^2}{L}\sum_{m\alpha}
\overline{G}^r_{\ell m}(\epsilon_\alpha-m\Omega)
[\overline{G}^r_{\ell' m}(\epsilon_\alpha-m\Omega)]^*
\nonumber \\
& & \times\delta(\omega-\epsilon_\alpha+m\Omega)
\langle \psi^\dagger_{m\alpha}\psi_{m\alpha}\rangle
\nonumber \\
& = & 2i\Gamma\sum_m
\overline{G}^r_{\ell m}(\omega)[\overline{G}^r_{\ell' m}(\omega)]^*
f(\omega+m\Omega), \label{eq:gwlss}
\end{eqnarray}
where each reservoir has its own Fermi energy shifted by $\ell\Omega$
and $\langle \psi^\dagger_{\ell\alpha}\psi_{\ell\alpha}\rangle
=f(\epsilon_\alpha)$. The above expression is quite appealing and
physically transparent. With the dissipation provided by the particle
reservoirs, all electron statistics are governed by the Fermi statistics
of the reservoirs and the effective tunneling between site $\ell$ and the reservoir
attached at site $m$ is given by the retarded Green's function
$\overline{G}^r_{\ell m}$. It is noted that we use the infinite-band
approximation for each fermion reservoirs so that any reservoir can
provide electrons to any other tight-binding lattice sites in principle, and
all possible thermal factors mix throughout the lattice.

Now, we turn to calculation of retarded Green's functions. With the
time-independent Hamiltonian, we only need to invert the matrix as
$\overline{G}^r_{\ell\ell'}(\omega)=[{\bf M}(\omega)^{-1}]_{\ell\ell'}$
with
\begin{equation}
[{\bf M}(\omega)]_{\ell\ell'}=(\omega+\ell\Omega+i\Gamma)\delta_{\ell\ell'}
+\gamma\delta_{|\ell-\ell'|,1},
\end{equation}
where the retarded self-energy $-i\Gamma$ is attached to each site
$\ell$ of the tight-binding lattice with the potential slope. Solution
to the matrix inversion can be found~\cite{aoki} as
\begin{equation}
\overline{G}^r_{\ell\ell'}(\omega)
=\sum_m\frac{J_{\ell-m}(\frac{2\gamma}{\Omega})J_{\ell'-m}(\frac{2\gamma}{\Omega})}{
\omega+m\Omega+i\Gamma},
\label{eq:grll}
\end{equation}
which can be easily verified from ${\bf M}(\omega)\overline{\bf
G}^r(\omega)={\bf I}$. Substituting this Green's function into
Eq.~(\ref{eq:gwlss}) gives the identical result Eq.~(\ref{eq:gtlss}) as
derived from the time-dependent temporal gauge.
Coming back to the retarded Green's function, we can easily confirm the
identity Eq.~(\ref{eq:gwr}) from a
straightforward calculation after substituting Eq.~(\ref{eq:grll}) into 
Eq.~(\ref{eq:gwr}) and by using the contour integral and the
completeness relation of Bessel functions.
From Eq.~(\ref{eq:grll}), it follows that
\begin{equation}
\overline{G}^r_{\ell+k,\ell'+k}(\omega)
=\overline{G}^r_{\ell\ell'}(\omega+k\Omega),
\end{equation}
from which Eq.~(\ref{eq:gwlss}) satisfies
\begin{equation}
\overline{G}^<_{\ell+k,\ell'+k}(\omega)
=\overline{G}^<_{\ell\ell'}(\omega+k\Omega).
\end{equation}
In an interacting model, the self-energy is expressed in $G^{r,<,>}$ and
inherits the same property,
\begin{equation}
\overline{\Sigma}^{r,<}_{\ell+k,\ell'+k}(\omega)
=\overline{\Sigma}^{r,<}_{\ell\ell'}(\omega+k\Omega).
\end{equation}

\subsection{Distribution function}

\begin{figure}
\rotatebox{0}{\resizebox{2.8in}{!}{\includegraphics{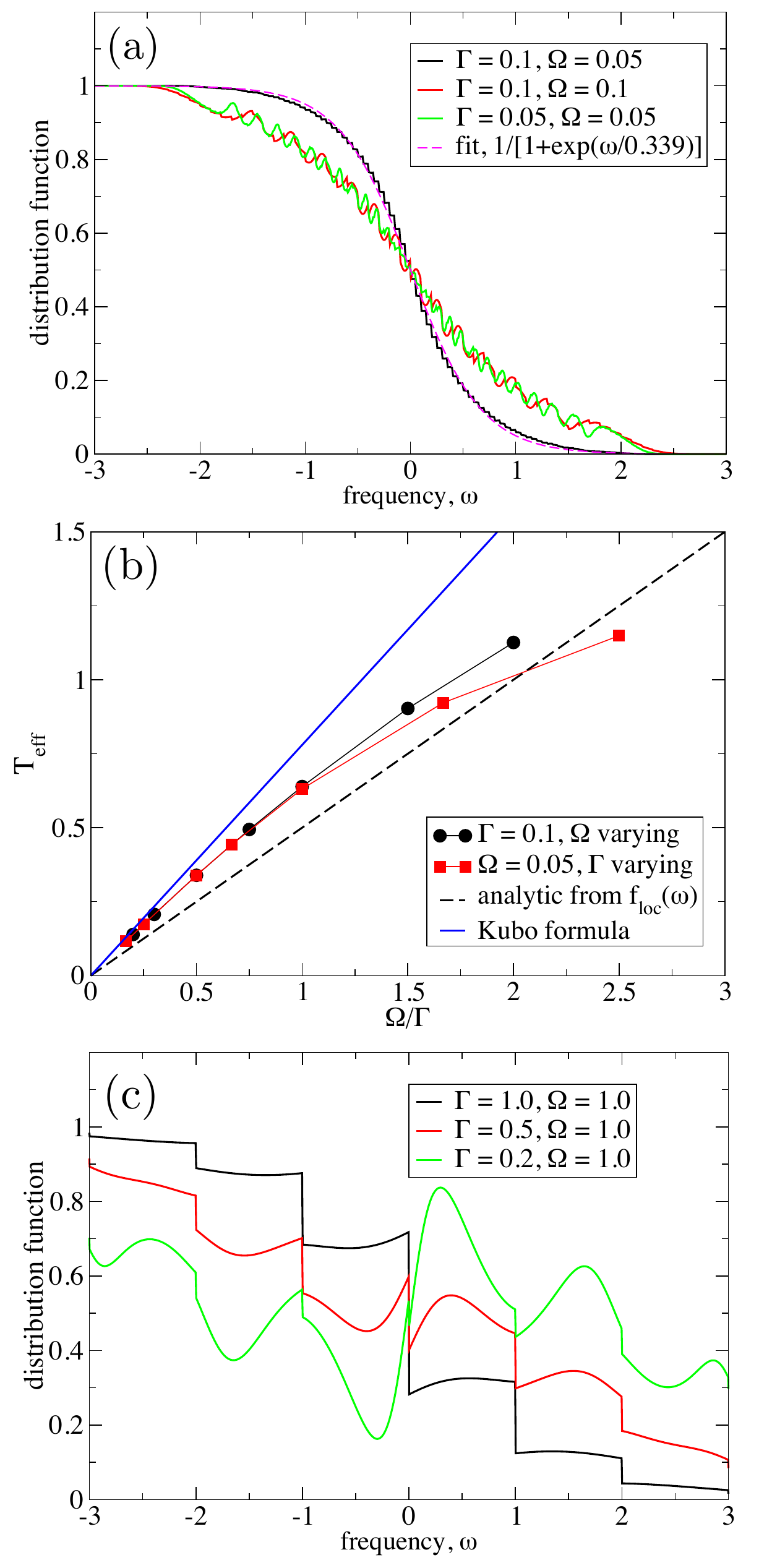}}}
\caption{(a) Local distribution function $f_{\rm loc}(\omega)$ for
several parameters of the damping $\Gamma$ and the Bloch oscillation
frequency $\Omega$ ($\Gamma,\Omega\ll\gamma$).  The effective temperature
is estimated through a fit to the Fermi-Dirac function. (b) The
effective temperature $T_{\rm eff}$ as a function of $\Omega/\Gamma$. Up to
$\Omega/\Gamma\approx 1$, $T_{\rm eff}$ is well described as a linear function
of $\Omega/\Gamma$. The dashed line denotes the analytic expression,
Eq.~(\ref{app:Teff}), derived from the low frequency $f_{\rm
loc}(\omega)$ as shown in Appendix A. The blue line is from the Kubo
formula with $T_{\rm eff}=(6/\pi^2)^{1/2}\gamma(\Omega/\Gamma)$. See Appendix B and discussions in section III.D. (c) For larger field
($\Omega=\gamma=1$), the steps become more prominent and the definition of
the effective temperature becomes less robust. Nevertheless, the trend in
(a-b) continues and at $\Gamma=0.2$ even a population inversion happens.
}
\label{fig3}
\end{figure}

The discussion so far has demonstrated explicitly that the dissipative
system with fermion baths can be described within the steady-state
formalism using the scattering state basis. One of the central
quantities to calculate is the effective local distribution function,
\begin{equation}
f_{\rm loc}(\omega)=-\frac{{\rm Im}\overline{G}^<_{00}(\omega)}{2
{\rm Im}\overline{G}^r_{00}(\omega)}
=\frac{\sum_\ell|\overline{G}^r_{0\ell}(\omega)|^2f(\omega+\ell\Omega)}{
\sum_\ell|\overline{G}^r_{0\ell}(\omega)|^2},
\label{eq:floc}
\end{equation}
where Eq.~(\ref{eq:gwr}) has been used for ${\rm
Im}\overline{G}^r_{00}(\omega)=-\Gamma\sum_\ell
|\overline{G}^r_{0\ell}(\omega)|^2$.
This result takes the same form as the ansatz considered in Aron \textit{et
al.}~\cite{aron2}.

FIG.~\ref{fig3}(a) shows the numerical evaluation of the local
distribution function $f_{\rm loc}(\omega)$. For $\Gamma,\Omega\ll\gamma$,
$f_{\rm loc}(\omega)$ is a superposition of small steps coming from the
thermal factors $f(\omega+\ell\Omega)$ in Eq.~(\ref{eq:floc}) with the
envelope following a smooth profile similar to the Fermi-Dirac function. Even
though there is no reason to expect that the nonequilibrium distribution
mimics the Fermi-Dirac function, we can nevertheless fit the result to
the function with an effective temperature $T_{\rm eff}$
as shown, despite some deviation (see Appendix A for more
details). As $T_{\rm eff}$ grows towards the finite tight-binding
bandwidth $4\gamma$, the fit becomes inaccurate.

The envelope of the local distribution function plays a similar role of
the Fermi-Dirac function which dictates the abundance of electron-hole
pairs available for interaction.
In the presence of additional many-body interactions such as the 
electron-phonon coupling to local optical phonons, the available
electron-hole pairs for inelastic dissipation are given by the $f_{\rm loc}(\omega)$
profile and the effective temperature $T_{\rm eff}$ in the Fermi-Dirac
function will play the role of hot electron temperature effectively.

It is remarkable that $T_{\rm eff}$ seems to approach infinity as the
damping parameter $\Gamma$ becomes smaller. Although it may look
counter-intuitive at first, this is only the manifestation of
the \textit{short-circuit} behavior where a finite voltage applied across a
low resistance conductor induces an extremely hot temperature. This is
also consistent with the numerical calculations with the general conclusion that
the electron temperature reaches an infinity in closed
interacting models.

Numerical fit indicates that $T_{\rm eff}$ is an increasing function
of $\Omega/\Gamma$ for a wide range of $(\Omega,\Gamma)$, although the
functional form eventually deviates from the form as shown in
FIG.~\ref{fig3}(b). For small $\Omega/\Gamma$, the effective temperature
behaves as
\begin{equation}
T_{\rm eff}\approx C\gamma\left(\frac{\Omega}{\Gamma}\right),
\label{eq:Teff}
\end{equation}
with a dimensionless numerical constant $C$. This equation is one of the
key results of this paper. In Appendix A, we derive the above
linear dependence of $(\Omega/\Gamma)$ and approximately estimate that
the constant $C\sim\frac12$ by analyzing the $\omega=0$ step in $f_{\rm
loc}(\omega)$. The effective temperature has been
previously observed in the momentum distribution function and has been
speculated~\cite{fbath} to behave as $T_{\rm eff}=\Omega^2/\Gamma$
based on the DC conductivity analogy. More careful and quantitative
analysis now shows that the correct dependence is the above relation,
Eq.~(\ref{eq:Teff}). The $T_{\rm eff}$ relation is even further
corroborated with $C$ derived from the Kubo formula (see Appendix B and
discussions in section III.D.) as the blue line in FIG.~\ref{fig3}(b).
The Kubo formula result 
\begin{equation}
T_{\rm
eff}=\left(\frac{6}{\pi^2}\right)^{1/2}\gamma\left(\frac{\Omega}{\Gamma}\right)
=0.7796\,\gamma\left(\frac{\Omega}{\Gamma}\right)
\label{eq:Teff2}
\end{equation} 
should be exact for
the limit $\Omega/\Gamma\to 0$.
The two analytical estimates bracket the numerical
$T_{\rm eff}$ [see FIG.~\ref{fig3}(b)], which shows that the expressions
Eqs.~(\ref{eq:Teff}) and (\ref{eq:Teff2}) are a
reliable approximation for $\Gamma$ and $\Omega$ up to $\gamma$.

The divergent effective temperature should
be taken with a caution to interpret in finite bandwidth systems.
Unlike with the quadratic
dispersion relation for continuum models~\cite{price}, the kinetic energy in the
single-band tight-binding model is always bounded and thus cannot give
off extremely \textit{hot}-electrons to the environment.

As $\Omega$ and $\Gamma$ become comparable to the bandwidth [see
FIG.~\ref{fig3}(c)], the
signature of Bloch oscillation steps become more obvious and the
definition of the effective temperature as determined by the shape of
the overall $f_{\rm loc}(\omega)$ is not very robust. While the
infinite temperature in a finite bandwidth system may be
questionable, the trend observed in FIG.~\ref{fig3}(a-b) continues.
In the small damping limit at $\Gamma=0.2$, even a population inversion
happened in the local distribution function.

\subsection{Time-evolution of wave-packet}

\begin{figure}
\rotatebox{0}{\resizebox{3.3in}{!}{\includegraphics{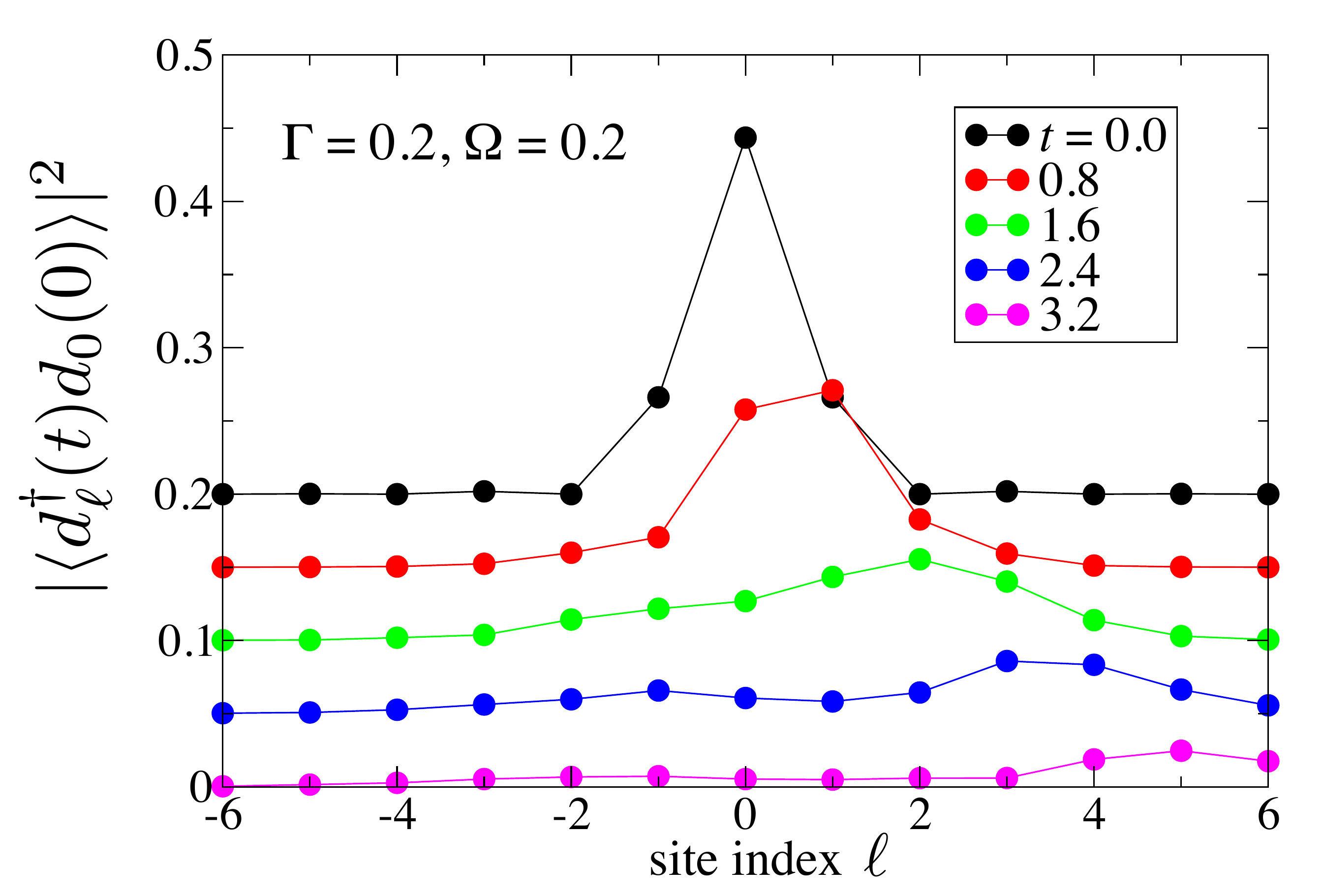}}}
\caption{Observation of wave-packet evolution out of nonequilibrium
steady-state. Disturbance created out of the steady-state travels down
the tight-binding ladder as time evolves. The amplitude of the
wave-packet diminishes due to the dephasing provided by the fermion
baths. The curves have been off-set for better visibility.
}
\label{fig4}
\end{figure}

So far, we have seen that the steady-state formalism provides a
convenient theoretical framework for nonequilibrium lattice model of
fermion baths. To better understand the nonequilibrium steady-state, we now look at
time-dependent quantity indicative of wave-packet drift. A steady-state
by definition is a time-independent reference state where 
a direct observation of time-evolution of a moving particle cannot be made. To
confirm that an electric charge moves down the potential slope as a
function of time, we create a hole out of occupied states~\cite{create}
at a central position and observe its movement to a different position
$x_h(\ell)$ as a function of time in
\begin{equation}
|\langle x_h(\ell),t|x_h(0),0\rangle|^2=|\langle
d^\dagger_\ell(t)d_0(0)\rangle|^2
=|G^<_{0\ell}(-t)|^2.
\end{equation}
The lesser Green's function can be easily decomposed in terms of
scattering states and
FIG.~\ref{fig4} shows the wave-packet traveling in the direction of the
applied field. Amplitude of the observable decays as $e^{-2\Gamma t}$
due to the dephasing of electrons from the fermion baths.

As depicted in Fig.~\ref{fig2}(b), electrons travel down the potential
slope in the Coulomb gauge by creating a trail of electron-hole (e-h)
pairs in the reservoirs. As long as the bandwidth of the reservoirs is
greater than the potential drop $\Omega$ between neighboring sites (we
assumed that the bandwidth is infinite in explicit calculations), each
reservoirs can accommodate an e-h pair with its energy matching $\Omega$
by particle exchange via tunneling. For narrower reservoir bandwidths,
multiple e-h pairs should be created to establish a DC current.  Since
the open reservoirs are of infinite length, the created e-h pairs travel
indefinitely inside the reservoirs and therefore the fermion baths
can produce effects similar to the inelastic processes in bosonic baths.
We discuss this further in the subsequent sections.

\subsection{Dissipation and energy flux}

We turn to discussions of energy dissipation. 
The Hamiltonian, Eq.~(\ref{eq:hcoul}), can be divided into three parts
as
\begin{equation}
\hat{H}_{\rm sys}=\hat{H}_{\rm TB}+\hat{H}_{\rm bath}+\hat{H}_{\rm
coup},
\end{equation}
with each term representing each line in Eq.~(\ref{eq:hcoul}),
respectively. In the steady-state limit, the energies stored in
$\hat{H}_{\rm TB}$ and $\hat{H}_{\rm coup}$ are stationary
$\frac{d}{dt}\langle \hat{H}_{\rm TB}\rangle = \frac{d}{dt}\langle
\hat{H}_{\rm coup}\rangle=0$, as will be demonstrated below. Unlike the
case with $\hat{H}_{\rm TB}$ and $\hat{H}_{\rm coup}$ which are of finite
spatial extent, the energy flux in $\hat{H}_{\rm bath}$ can be non-zero.
In the scattering theory~\cite{doyon}, the scattering states are
formulated in the limit that the spatial extent of the scattered wave
($L_{\rm scatt}$) into the reservoirs is much shorter than the length of the reservoir chain
($L_{\rm scatt}\ll L$). Therefore the scattering state represents a
solution that the scattered wave \textit{constantly propagating inside
the reservoirs} without being backscattered from the edge of the
reservoirs, and quantities involving the extended states
$c^\dagger_{\ell k}$ or $c_{\ell k}$ do not have to be stationary in
general~\cite{stationary}. 

To be more concrete, we discuss explicit calculations. With
the fermion baths, a DC electric-field establishes a DC current $J$, as
calculated in Ref.~\onlinecite{fbath}. To investigate the effect of the
Joule heating, we consider $\hat{H}_{\rm TB}$
\begin{eqnarray}
\frac{d}{dt}\langle \hat{H}_{\rm TB}\rangle & = &
i\left\langle [\hat{H}_{\rm Coul},\hat{H}_{\rm TB}] \right\rangle \\
&=&\Omega\langle \hat{J}\rangle 
+i\gamma g\sum_\ell\langle
(\bar{c}^\dagger_{\ell+1}+\bar{c}^\dagger_{\ell-1})d_\ell
-{\rm H.c.} \rangle\nonumber
\end{eqnarray}
with the current operator
$\hat{J}=i\gamma\sum_\ell(d^\dagger_{\ell+1}d_\ell
-d^\dagger_\ell d_{\ell+1})$ within the main chain and 
$\bar{c}_\ell=(1/\sqrt{L})\sum_\alpha
c_{\ell\alpha}$. We used the steady-state condition for the occupation
$\frac{d}{dt}\langle d^\dagger_\ell d_\ell\rangle =0$.
The first term represents the Joule heating and the
second term the energy flux of electrons from the kinetic energy of the
main chain into the
coupling $\hat{H}_{\rm coup}$. Denoting the energy flux per site as $\hat{P}$, we show that
$\langle\hat{P}\rangle = -\Omega\langle\hat{J}\rangle$. We present the detailed
calculations in Appendix C and show the equality of the above equation
based on the scattering-state formalism.

It can be shown further that $\frac{d}{dt}\langle \hat{H}_{\rm
coup}\rangle=0$. As shown in appendix B, the energy influx to each of
the reservoirs is nothing but the Joule heating
\begin{eqnarray}
\frac{d}{dt}\langle \hat{h}_{\rm 
bath}\rangle & = & 2\Gamma\int \omega A_{\rm loc}(\omega)[f_{\rm loc}(\omega)
-f(\omega)]d\omega\nonumber \\
&= &\Omega\langle J\rangle,
\label{joule}
\end{eqnarray}
with the local spectral function defined as
\begin{equation}
A_{\rm loc}(\omega)=-\frac{1}{\pi}{\rm Im}G^r_{00}(\omega).
\end{equation}
The lowercase Hamiltonian $\hat{h}$ denotes the corresponding
Hamiltonian per tight-binding site.
It might sound paradoxical that the energy in the electronic
system is non stationary,
\begin{equation}
\frac{d}{dt}\langle \hat{h}_{\rm sys}\rangle=
\frac{d}{dt}\langle \hat{h}_{\rm bath}\rangle=\Omega\langle J\rangle
\approx\frac{4\gamma\Gamma\Omega^2}{\pi(\Omega^2+4\Gamma^2)}.
\end{equation}
(Here the last equality is from the steady-state current taken from
Ref.~\onlinecite{fbath}.)
This is due to the fact that, although $\hat{H}_{\rm sys}$ governs the
electron dynamics, there is another part of Hamiltonian which should be
included for a closed system -- the battery connected across the
tight-binding chain. Since the battery loses its
stored charge $Q$ with the rate of $\dot{Q}=-\langle J\rangle$, the electrostatic 
energy decrease per
unit cell of the tight-binding chain becomes $\frac{d}{dt}\langle
\hat{h}_{\rm battery}\rangle =-\Omega\langle J\rangle$, and the total
energy $\hat{H}_{\rm tot}=\hat{H}_{\rm sys}+\hat{H}_{\rm battery}$ is
stationary in the steady-state.

The discussion here again confirms the picture depicted in FIG.~\ref{fig2}(b) where the
fermion baths act as energy reservoirs while the net electron number
flux into the reservoirs is zero. Despite their simplicity, the fermion baths
through their particle-hole excitations play the role of bosonic baths,
apart from the boson's explicit dispersion relation (with the exception
of the Luttinger liquid bath) and the physics that might occur
from the nonlinear effect of the bosonic statistics.

\subsection{Steady-state current for interacting systems}

From the energy dissipation relations above, we obtain the useful
formula for the steady-state current,
\begin{equation}
\langle\hat{J}\rangle=\frac{2\Gamma}{\Omega}\int\omega A_{\rm loc}(\omega)[f_{\rm
loc}(\omega)-f(\omega)]d\omega,
\label{eq:current}
\end{equation}
where only on-site Green's functions are needed as in Meir-Wingreen formula in
quantum dot transport~\cite{meir}. To recover the Ohm's law for small
field that $\langle \hat{J}\rangle\propto \Omega$ one should have that
the integral goes as $\Omega^2$ as the
leading order. This is justified since applying a field of opposite
direction $-\Omega$ should not change the local properties and the
integral should be of order $\Omega^2$. This argument can be used to
analyze the linear response limit, as described below.

The above relation Eq.~(\ref{eq:current}), verified explicitly for
the non-interacting model in Appendix C, can be extended to interacting models. The
key identities are steady-state conditions
\begin{equation}
\frac{d}{dt}\langle d^\dagger_{\ell\sigma}d_{\ell\sigma}\rangle =
0\mbox{ and }
\frac{d}{dt}\langle \hat{h}_{\rm bath}\rangle =
\Omega\langle\hat{J}\rangle,
\end{equation}
which we expect to hold generally for interacting systems as long as the
interaction potential does not hold infinite amount of energy per site,
as in Hubbard model. Here we used the spin index $\sigma$.
With on-site interaction,
$\langle\dot{n}_{d\sigma}\rangle=0$ ensures zero particle flux into the
baths. The energy
flux equation and the Dyson'e equation hold as Eq.~(\ref{app:bath})
and (\ref{app:dyson}), respectively. This immediately shows that the
current,~(\ref{eq:current}), holds for a wide range of interacting
fermion bath models.

Steady-state current derived by Meir-Wingreen~\cite{meir} has been
widely used in quantum dot calculations. The formula~(\ref{eq:current})
can be seen as its extension for lattice models with fermion baths. The
equation is a functional of only local Green's functions. However, it
should be made clear that while
the simplified Meir-Wingreen formula for a single quantum
model only requires $G^{\rm r}_{\rm QD}(\omega)$ for the quantum dot, both
of $G^r_{\rm loc}(\omega)$ and $G^<_{\rm loc}(\omega)$ are
necessary in the lattice models.

Using the key equation (\ref{eq:current}), and Eq.~(\ref{eq:Teff}) in
the non-interacting limit, a linear response limit can be analyzed. In the limit of
$\Omega/\Gamma\to 0$, the effective temperature is expected to be
small and, therefore, we can use the Sommerfeld expansion~\cite{ashcroft}
to derive the linear electrical current
\begin{equation}
J_0=e\frac{2\Gamma}{\Omega}\frac{\pi^2}{6}T_{\rm eff}^2A_{\rm loc}(0)
=eC^2\frac{\pi^2}{3}\frac{\gamma^2}{\Gamma}A_{\rm loc}(0)\cdot\Omega,
\end{equation}
and we obtain the linear DC conductivity 
\begin{equation}
\sigma_0=J_0/E=C^2\frac{\pi^2}{3} e^2a\gamma^2 A_{\rm loc}(0)\cdot
\Gamma^{-1}.
\label{eq:cond}
\end{equation}
Here $A_{\rm loc}(0)$ is the equilibrium spectral function evaluated at
the Fermi energy. With $\gamma A_{\rm loc}(0)\sim 1$, $\gamma\sim
{m^*}^{-1}$ and the scattering time $\tau\sim\Gamma^{-1}$, we recover
the Drude conductivity~\cite{fbath}. Comparing this with the linear
response theory using the Kubo formula, we obtain
$\sigma_0=2e^2a\gamma^2/(\pi\Gamma\sqrt{\Gamma^2+4\gamma^2})$ as
detailed in Appendix B. Noting that $A_{\rm
loc}(0)=(\pi\sqrt{\Gamma^2+4\gamma^2})^{-1}$, we have from
Eq.~(\ref{eq:cond}) that $C=\sqrt{6/\pi^2}=0.7796$. This result is shown
as the blue line in FIG.~\ref{fig3}(b) in comparison to the numerically
obtained $T_{\rm eff}$.

In the interacting limit, the
effective temperature expression Eq.~(\ref{eq:Teff}) should be modified.
We expect that the same form $T_{\rm
eff}=C^*\gamma^*(\Omega/\Gamma^*)$ holds with renormalized parameters
$C^*$, $\gamma^*$ and $\Gamma^*$. Then the linear response equation 
becomes $\sigma_0={C^*}^2\frac{\pi^2}{3} e^2a{\gamma^*}^2 A_{\rm
loc}(0)\cdot (\Gamma/{\Gamma^*}^2)$.

\section{conclusion}

In this work, we have reformulated the electron transport in
tight-binding lattice driven by a DC electric field using both
time-dependent and time-independent gauges. The time-independent Coulomb gauge
with fermion baths leads to the scattering state description for
steady-state, which makes the calculation and interpretation more
intuitive. Nonequilibrium quantum statistics of quantum dot model, as
proposed by Hershfield~\cite{hershfield}, can be extended to
nonequilibrium lattice as summarized in the scattering state
expressions,
\begin{eqnarray}
\hat\rho_{\rm noneq} & = & \exp\left[
-\beta_{\rm
bath}\sum_{\ell\alpha}\epsilon_\alpha\psi^\dagger_{\ell\alpha}
\psi_{\ell\alpha}
\right]\\
\hat{H} & = & \sum_{\ell\alpha}(\epsilon_\alpha-\ell\Omega)\psi^\dagger_{\ell\alpha}
\psi_{\ell\alpha},
\end{eqnarray}
with the inverse temperature $\beta_{\rm bath}$ of the baths. The
reservoir scattering states (represented by $\psi_{\ell\alpha}$)
are shifted by the applied electrostatic potential
$\ell\Omega$, and the chemical potential is simultaneously shifted with the
electrostatic potential. Therefore, the energy spectra governing the dynamics and
statistics are different in the above expression.
The formalism provides a natural framework for approximations such as
the dynamical mean-field theory (DMFT).

It has been shown that the fermion bath model, although quite
rudimentary, produces dissipation mechanism consistent with the
Boltzmann transport theory. In particular, the steady-state effective
temperature induced by the external field depends quite strongly on the
electric field and the damping. The effective temperature becomes
divergent as $T_{\rm eff}\propto\Omega/\Gamma$
for small damping $\Gamma$ versus the Bloch frequency $\Omega=eEa$ ($a$
is the lattice constant, and $E$ the electric field). 
Although this might look surprising at first,
this phenomenon is simply the manifestation of the short-circuit effect.
It also verifies various numerical calculations with the infinite electron
temperature resulting in isolated lattice models. These findings have fundamental
implications in nonequilibrium quantum statistics in that dissipation
processes cannot be implicitly included as thermalization as in the
Boltzmann factor of equilibrium Gibbsian statistics. Through the energy
dissipation and the
Joule heating in the fermion reservoirs, a general DC current relation
Eq.~(\ref{eq:current})
has been derived for interacting models, as an extension of the
Meir-Wingreen formula to nonequilibrium lattice systems. The linear
response limit has been confirmed within this formalism.

Despite the lack of momentum scattering and explicit inelastic processes,
the generic features of the fermion bath model which are consistent with
semi-classical theory are quite significant. Furthermore, for its
simplicity the fermion bath model can be used as an ideal building block for studying
strong correlation effects in lattice driven out of equilibrium.
Particularly, with the time-independent Coulomb gauge DMFT can be
readily formulated using the scattering state
method~\cite{prl07,anders,prb06,aron2} It is well-known in equilibrium
strong correlation physics that electrons undergo collective state when
a strong interaction is present, with some emergent energy scale $T^*$.
One may speculate that an electric field of order $\Omega\sim T^*$ would
significantly alter the strongly correlated state. However, our study
suggests that the dissipation strongly interplays with the nonequilibrium
condition and non-trivial physics may arise even at $\Gamma<\Omega\ll
T^*$. Further systematic studies are necessary to understand the
interplay of nonequilibrium and strong correlation effects.

\section{acknowledgement}
We thank helpful discussions with Kwon Park, Woo-Ram Lee, Jainendra
Jain, Anthony Leggett, Natan Andrei and Gabi
Kotliar. This work has been financially supported by the National
Science Foundation through Grant No. DMR- 0907150.

\appendix

\section{Analytic estimate of effective temperature from $f_{\rm
loc}(\omega)$}

\begin{figure}
\rotatebox{0}{\resizebox{3.3in}{!}{\includegraphics{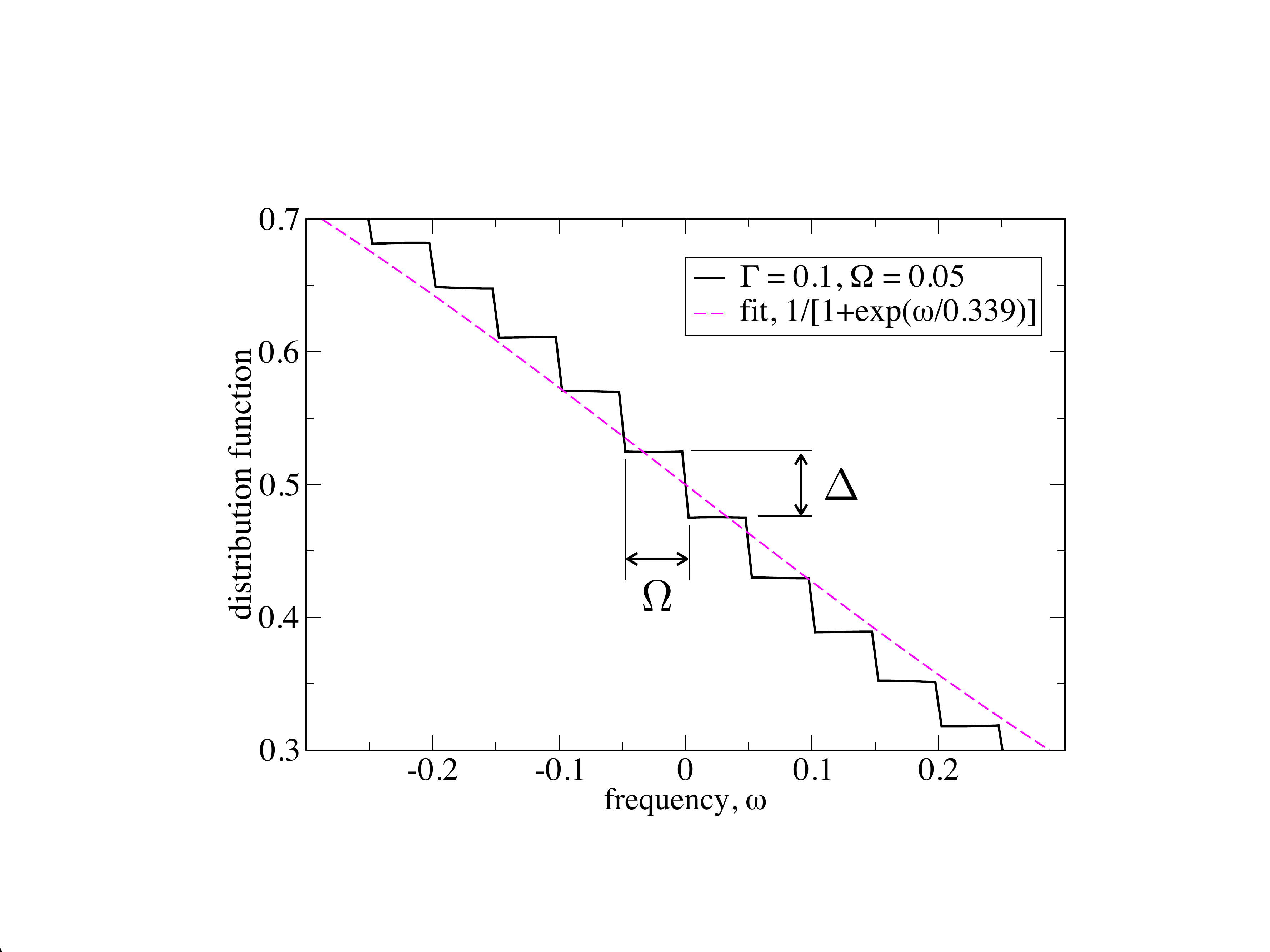}}}
\caption{Fit of $f_{\rm loc}(\omega)$ at low frequency $\omega$. An
analytic expression of the effective temperature $T_{\rm eff}$ is
estimated from the low frequency part plotted in FIG. 3. The slope of
the fit is approximated from the step $\Delta$ at $\omega=0$ as $-\Delta/\Omega$.
The actual fit gives a somewhat smaller slope than the analytic estimate.
}
\label{fig5}
\end{figure}

Here we analytically justify the relation $T_{\rm eff}\propto
\gamma(\Omega/\Gamma)$ by considering the low frequency steps as shown in
FIG.~\ref{fig5}. The first step $\Delta$ at $\omega=0$ can be expressed as
\begin{equation}
\Delta=-\frac{\Gamma|\overline{G}^r_{00}(0)|^2}{{\rm
Im}\overline{G}^r_{00}(0)}=\Gamma\left\{
{\rm Im}\left[\overline{G}^r_{00}(0)^{-1}\right]
\right\}^{-1}.
\end{equation}
Here we look at the limit of small $\Omega$ and approximate
$\overline{G}^r_{00}(0)$ by the equilibrium Green function
\begin{equation}
\overline{G}^r_{00}(\omega)^{-1}\approx (\omega+i\Gamma)\left[
1-\frac{4\gamma^2}{(\omega+i\Gamma)^2}
\right]^{1/2}.
\end{equation}
Therefore, the analytic expression for the slope of the fit becomes
\begin{equation}
-\frac{\Delta}{\Omega}=-\frac{\Gamma}{\Omega\sqrt{4\gamma^2+\Gamma^2}}
\approx -\frac{\Gamma}{2\gamma\Omega}.
\end{equation}
By equating this to the slope of the effective Fermi-Dirac function
$(1+e^{\omega/T_{\rm eff}})^{-1}$, we obtain
\begin{equation}
T_{\rm eff}\approx\frac{\gamma}{2}\left(
\frac{\Omega}{\Gamma}\right).
\label{app:Teff}
\end{equation}
Note that while the actual numerical fit overestimates $T_{\rm eff}$
from the analytic expression
due to the high-frequency contribution, the overall functional
dependence is quite reliable for $\Gamma,\Omega<\gamma$.

\section{Conductivity from linear response theory}

From the Kubo formula~\cite{mahan}, the linear conductivity can be
exactly calculated in the small $\Omega$ limit. For convenience, we
calculate the current-current correlation function in the imaginary-time
formalism and then analytically
continue to the real-frequency in the optical
conductivity~\cite{pruschke}. For the uniform (${\rm q}=0$) response
function in the Matsubara frequency $i\nu$, the conductivity is
expressed as
\begin{equation}
\sigma(i\nu)=\frac{i}{i\nu}\frac{1}{L\beta}\sum_{k,n}
v_k^2G_k(i\omega_n)G_k(i\omega_n+i\nu).
\end{equation}
Here $v_k=2\gamma\sin(k)$ is the group velocity and the Matsubara Green's
function for the electron is given as
\begin{equation}
G_k(i\omega_n)=\frac{1}{i\omega_n-\epsilon_k+i\Gamma(\omega_n/|\omega_n|)}
=\int d\epsilon\frac{\rho_0(\epsilon-\epsilon_k)}{i\omega_n-\epsilon},
\end{equation}
with $\rho_0(\epsilon)=\Gamma/\pi\cdot(\epsilon^2+\Gamma^2)^{-1}$.
Performing the Matsubara summation and then the analytic continuation
$i\nu\to\omega+i\eta$ for finite $\omega$, we have
\begin{eqnarray}
\sigma(\omega)&=&\frac{i}{L\omega}\sum_kv_k^2\int d\epsilon_1\int
d\epsilon_2\rho_0(\epsilon_1-\epsilon_k)\rho_0(\epsilon_2-\epsilon_k)\nonumber
\\
& & \times \frac{f(\epsilon_1)-f(\epsilon_2)}{\omega+\epsilon_1-\epsilon_2+i\eta}.
\end{eqnarray}
Taking its real part and the static limit $\omega\to 0$ at zero
temperature, we obtain the DC linear conductivity
\begin{eqnarray}
\sigma_0&=&\frac{4e^2a\gamma^2\Gamma^2}{\pi}\int_0^{2\pi}\frac{dk}{2\pi}\frac{\sin^2
k}{(\Gamma^2+4\gamma^2\cos^2k)^2}\nonumber \\
&=&\frac{2e^2 a\gamma^2}{\pi\Gamma\sqrt{\Gamma^2+4\gamma^2}},
\end{eqnarray}
with the restored constants $e$ and $a$.

\section{Joule heating and energy flux}
The current expectation value measured at the site $\ell=0$
is expressed as $\langle \hat{J}\rangle=2\gamma{\rm Im}\langle
d^\dagger_1 d_0\rangle$. Using the scattering state basis, we have
\begin{eqnarray}
\langle d^\dagger_1 d_0\rangle
& = &
\frac{g^2}{L}\sum_{n\alpha}
[\overline{G}^r_{1n}(\epsilon_\alpha-n\Omega)]^*
\overline{G}^r_{0n}(\epsilon_\alpha-n\Omega)f(\epsilon_\alpha)
\nonumber \\
& = &
\frac{\Gamma}{\pi}\sum_{nkk'}\int_{-\infty}^0 d\omega
\frac{J_{1-k-n}J_{-k}J_{-k'-n}J_{-k'}}{(\omega+k\Omega-i\Gamma)
(\omega+k'\Omega+i\Gamma)} \nonumber \\
& = &
-\frac{\Gamma}{\pi}\sum_{k}\int_{-\infty}^0 
\frac{J_{k}J_{k-1}d\omega
}{(\omega+k\Omega-i\Gamma)
[\omega+(k-1)\Omega+i\Gamma]},\nonumber
\end{eqnarray}
by using Eq.~(\ref{eq:grll}) and the Bessel function identities. We have
suppressed the argument in the Bessel functions. With
integration of elementary functions we obtain
\begin{eqnarray}
\langle d^\dagger_1 d_0\rangle
& = & \frac{\Gamma}{\pi(\Omega-2i\Gamma)}\sum_k
J_k\left(\frac{2\gamma}{\Omega}\right)
J_{k-1}\left(\frac{2\gamma}{\Omega}\right)\nonumber \\
& &\times
\left[
\frac12\ln\frac{k^2\Omega^2+\Gamma^2}{(k-1)^2\Omega^2+\Gamma^2}
+i\chi_{k,k-1}
\right],
\end{eqnarray}
with $\chi_{nm}=\pi+\tan^{-1}\frac{n\Omega}{\Gamma}
+\tan^{-1}\frac{m\Omega}{\Gamma}$. This immediately confirms that the
current evaluated from the scattering state basis matches the result
in Ref.~\onlinecite{fbath}.

For the operator $\langle \hat{P}\rangle$ we need to calculate 
$\langle (\bar{c}^\dagger_1+\bar{c}^\dagger_{-1})d_0\rangle$.
The $\bar{c}^\dagger$-operators are expressed with the scattering state
basis as
\begin{equation}
\bar{c}^\dagger_\ell
=
\frac{1}{\sqrt{L}}\sum_{\alpha}\left[\psi^\dagger_{\ell\alpha}
+\frac{g^2}{L}\sum_{\ell'\alpha'}
\frac{[\overline{G}^r_{\ell\ell'}(\epsilon_{\alpha'}-\ell'\Omega)]^*
\psi^\dagger_{\ell'\alpha'}}{
\epsilon_{\alpha'}-\epsilon_\alpha-(\ell'-\ell)\Omega-i\eta}\right].
\end{equation}
A lengthy but straightforward calculation gives
\begin{equation}
{\rm Im}\langle (\bar{c}^\dagger_1+\bar{c}^\dagger_{-1})d_0\rangle
=\frac{\Omega}{2g\gamma}\langle\hat{J}\rangle
\label{app:cd0}
\end{equation}
which confirms the identity $\langle \hat{P}\rangle =-\Omega\langle
\hat{J}\rangle$.

For the energy flux into the fermion baths, we examine
\begin{eqnarray}
\langle \dot{\hat{H}}_{\rm bath}\rangle
&=& i\langle [\hat{H}_{\rm sys},\hat{H}_{\rm bath}]\rangle\nonumber \\
&= &
\frac{ig}{\sqrt{L}}\sum_{\ell\alpha}(\epsilon_\alpha-\ell\Omega)\langle
c^\dagger_{\ell\alpha}d_\ell-d^\dagger_\ell c_{\ell\alpha}\rangle.
\label{app:bath}
\end{eqnarray}
First we show that $\sqrt{L^{-1}}\sum_{\alpha}\langle
c^\dagger_{\ell\alpha}d_\ell-d^\dagger_\ell c_{\ell\alpha}\rangle=0$.
From the steady-state condition of $\frac{d}{dt}\langle d^\dagger_\ell
d_\ell\rangle = 0$, we have $\langle[\hat{H}_{\rm Coul},d^\dagger_\ell 
d_\ell]\rangle = -\gamma\langle d^\dagger_{\ell +1}d_\ell
-d^\dagger_\ell d_{\ell +1}+d^\dagger_{\ell -1}d_\ell 
-d^\dagger_\ell d_{\ell -1}
\rangle -\frac{g}{\sqrt{L}}\sum_\alpha \langle
c^\dagger_{\ell\alpha}d_\ell - d^\dagger_\ell c_{\ell\alpha}\rangle =
0$. The first term is the total flux into the $\ell$-th site due to the
current along the TB chain, and in the steady-state it is zero.
Therefore we have zero particle-flux into the reservoir,
$\sqrt{L^{-1}}\sum_{\alpha}\langle
c^\dagger_{\ell\alpha}d_\ell-d^\dagger_\ell c_{\ell\alpha}\rangle=0$.
The remaining summation $\sqrt{L^{-1}}\sum_{\alpha}\epsilon_\alpha\langle
c^\dagger_{\ell\alpha}d_\ell-d^\dagger_\ell c_{\ell\alpha}\rangle$ is
the energy flux measured with respect to the $\ell$-th reservoir
chemical potential level and it should be independent of $\ell$. Setting
$\ell=0$, we can rewrite the expression as follows.

Consider $G^<_{d\alpha}(t)=i\langle
c^\dagger_{0\alpha}(0)d_{0}(t)\rangle$
and $G^<_{\alpha d}(t)=i\langle
d^\dagger_0(-t)c_{0\alpha}(0)\rangle$. 
For the energy flux per reservoir, we can write
\begin{eqnarray}
\langle \dot{\hat{h}}_{\rm bath}\rangle
& = & \frac{g}{\sqrt{L}}\sum_\alpha \epsilon_\alpha \int [G^<_{d\alpha
}(\omega)-G^<_{\alpha d
}(\omega)]\frac{d\omega}{2\pi}.\nonumber
\end{eqnarray}
From the Dyson's equation~\cite{meir},
\begin{eqnarray}
&& G^<_{d\alpha }(\omega)-G^<_{\alpha d}(\omega)\nonumber \\
&& =\frac{-2\pi i g}{\sqrt{L}}\delta(\omega-\epsilon_\alpha)
\{G^<_{00}(\omega)-f(\omega)[G^a_{00}(\omega)-G^r_{00}(\omega)]\}
\nonumber \\
&& = \frac{4\pi^2 g}{\sqrt{L}}\delta(\omega-\epsilon_\alpha)
A_{\rm loc}(\omega)[f_{\rm loc}(\omega)-f(\omega)].
\label{app:dyson}
\end{eqnarray}
Then
\begin{equation}
\langle \dot{\hat{h}}_{\rm bath}\rangle
=2\Gamma\int\omega A_{\rm loc}(\omega)[f_{\rm
loc}(\omega)-f(\omega)]d\omega.
\end{equation}

Taking the time-derivative of the above Green's functions at $t=0$, we have from
Eq.~(\ref{app:dyson}),
\begin{eqnarray}
&& \left.\frac{g}{\sqrt{L}}\sum_\alpha
i\frac{d}{dt}\left[G^<_{d\alpha }(t)-G^<_{\alpha d}(t)\right]\right|_{t=0} 
\nonumber \\
&& =2\Gamma\int\omega A_{\rm loc}(\omega)[f_{\rm
loc}(\omega)-f(\omega)]d\omega=\langle \dot{\hat{H}}_{\rm
bath}\rangle.\nonumber
\end{eqnarray}
This can be equated to
\begin{eqnarray}
&&\frac{-ig}{\sqrt{L}} \sum_\alpha\langle
c^\dagger_{0\alpha}[\hat{H}_{\rm Coul},d_0]+[\hat{H}_{\rm
Coul},d^\dagger_0]c_{0\alpha}\rangle \nonumber \\
&=&
-ig\gamma\langle
\bar{c}^\dagger_0(d_{1}+d_{-1})-(d^\dagger_1+d^\dagger_{-1})\bar{c}_0\rangle
\nonumber \\
& = &
-ig\gamma\langle
(\bar{c}^\dagger_1+\bar{c}^\dagger_{-1})d_0-d^\dagger_0(\bar{c}_1+\bar{c}_{-1})\rangle,
\end{eqnarray}
where the translational invariance of the steady-state has been used.
Using Eq.~(\ref{app:cd0}), Eq.~(\ref{joule}) is confirmed.


\begin{thebibliography}{*}

\bibitem{kadanoff} Leo P. Kadanoff and Gordon Baym, \textit{Quantum
Statistical Mechanics}, Westview Press (1994).

\bibitem{mahan} G. D. Mahan, \textit{Many-Particle Physics} 3rd Ed.,
Chap. 8, Kluwer Academic (2000).

\bibitem{landauer} Yoseph Imry and Rolf Landauer, Rev. Mod. Phys. {\bf
71}, S306 (1999).

\bibitem{doyon} B. Doyon and N.Andrei, Phys. Rev. B {\bf 73}, 245326 (2006).

\bibitem{tdnrg} F. B. Anders and A. Schiller, Phys. Rev. Lett. {\bf 95}, 196801
(2005).

\bibitem{werner} P. Werner, T. Oka, and A.J. Millis, Phys. Rev. B
{\bf 79}, 035320 (2009).

\bibitem{schiro} Marco Schiro and Michele Fabrizio, Phys. Rev. B
{\bf 79}, 153302 (2009).

\bibitem{zubarev} D. N. Zubarev, \textit{Nonequilibrium Statistical
Thermodynamics} (Consultants Bureau, New York, 1974).

\bibitem{hershfield} S. Hershfield, Phys. Rev. Lett. {\bf 70}, 2134 (1993).

\bibitem{schiller} A. Schiller and S. Hershfield, Phys. Rev. B {\bf 51}, 12896
(1995).

\bibitem{mehta} P. Mehta and N. Andrei, Phys. Rev. Lett. {\bf 96},
216802 (2006).

\bibitem{prl07} J. E. Han and R. J. Heary, Phys. Rev. Lett. {\bf 99},
236808 (2007).

\bibitem{anders} F. B. Anders, Phys. Rev. Lett. {\bf 101}, 066804 (2008).

\bibitem{prb06} J. E. Han, Phys. Rev. B {\bf 73}, 125319 (2006); J. E.
Han, Phys. Rev. B {\bf 75}, 125122 (2007).

\bibitem{freericks} J. K. Freericks, Phys. Rev. B {\bf 77}, 075109 (2008).

\bibitem{dmft} A. Georges, G. Kotliar, W. Krauth, and M. J. Rozenberg,
Rev. Mod. Phys. {\bf 68}, 13 (1996).

\bibitem{turkowski} V. Turkowski and J. K. Freericks, Phys. Rev. B {\bf
71}, 085104 (2005).

\bibitem{eckstein} Martin Eckstein, Takashi Oka, and Philipp Werner,
Phys. Rev. Lett. {\bf 105}, 146404 (2010).

\bibitem{aoki} Naoto Tsuji, Takashi Oka, and Hideo Aoki, Phys. Rev. B
{\bf 78}, 235124 (2008); Naoto Tsuji, Takashi Oka, and Hideo Aoki, 
Phys. Rev. Lett. {\bf 103}, 047403 (2009).

\bibitem{demler} Takuya Kitagawa, Erez Berg, Mark Rudner, and Eugene
Demler, Phys. Rev. B {\bf 82}, 235114 (2010).

\bibitem{amaricci} A. Amaricci, C. Weber, M. Capone, and G. Kotliar,
Phys. Rev. B {\bf 86}, 085110 (2012).

\bibitem{aron} Camille Aron, Gabriel Kotliar, and Cedric Weber, Phys.
Rev. Lett. {\bf 108}, 086401 (2012).

\bibitem{aron2} Camille Aron, Cedric Weber, and Gabriel Kotliar,
Phys. Rev. B {\bf 87}, 125113 (2013).

\bibitem{vidmar} M. Mierzejewski, L. Vidmar, J. Bonca, and P. Prelovsek,
Phys. Rev. Lett. {\bf 106}, 196401 (2011); L. Vidmar, J. Bonca, T.
Tohyama, and S. Maekawa, Phys. Rev. Lett. {\bf 107}, 246404 (2011).

\bibitem{oka} Takashi Oka, and Hideo Aoki, Phys. Rev. Lett. {\bf 95},
137601  (2005).

\bibitem{fbath} Jong E. Han, Phys. Rev. B {\bf 87}, 085119 (2013).

\bibitem{lebwohl} Paul A. Lebwohl and Raphael Tsu, J. Appl. Phys. {\bf
41}, 2664 (1970).

\bibitem{meir} Y. Meir and N. S. Wingreen, Phys. Rev. Lett. {\bf 68}, 2512
(1992).

\bibitem{jauho} Antti-Pekka Jauho, Ned S. Wingreen and Yigal Meir, Phys.
Rev. B {\bf 50}, 5528 (1994).

\bibitem{blandin} A. Blandin, A. Nourtier, D. W. Hone, J. Phys.
(Paris) {\bf 37}, 369 (1976).

\bibitem{gradshteyn} I. S. Gradshteyn and I. M. Rhizyk, \textit{Table of
Integrals, Series, and Products}, formulas 8.452, 8.453, and 8.530, 7th Ed.
Elsevier (2007).

\bibitem{scattering} M. Gell-Mann and M. L. Goldberger, Phys. Rev. {\bf
91}, 398 (1953).

\bibitem{price} Peter J. Price, J. Appl. Phys. {\bf 53}, 6863 (1982).

\bibitem{create} The sum of group velocity of occupied and empty states
out of a closed band is zero. Therefore, if we create an electron into
empty states, the disturbance travels in the opposite direction.

\bibitem{stationary} Although $\hat{H}_{\rm coup}$ contains
$c_{\ell k}$, $\sqrt{L^{-1}}\sum_k c_{\ell k}$ represents the first
orbital in the reservoir that couples to the tight-binding chain.
Therefore $\hat{H}_{\rm coup}$ is a local operator.

\bibitem{ashcroft} N. W. Ashcroft and N. D. Mermin, \textit{Solid State
Physics}, Thomson Learning (1976).

\bibitem{pruschke} Th. Pruschke, D. L. Cox, and M. Jarrell, Phys. Rev. B
{\bf 47}, 3553 (1993).













\end{thebibliography}
\end{document}